\documentclass{kluwer}    

\newdisplay{guess}{Conjecture}
\usepackage{graphicx}

\begin{document}
\begin{article}
\begin{opening}         
\title{Masses and Radii of Low-Mass Stars: Theory versus Observations}
\author{I. \surname{Ribas}}
\runningauthor{Ribas}
\runningtitle{Masses and Radii of Low-Mass Stars}
\institute{
Institut de Ci\`encies de l'Espai -- CSIC, Campus UAB,
Facultat de Ci\`encies, Torre C5 - parell - 2a planta, 08193 Bellaterra,
Spain; e-mail: iribas@ieec.uab.es\\
Institut d'Estudis Espacials de Catalunya (IEEC), Edif.
Nexus, C/Gran Capit\`a, 2-4, 08034 Barcelona, Spain
}
\date{September 28, 2005}

\begin{abstract}
Eclipsing binaries with M-type components are still rare objects. Strong
observational biases have made that today only a few eclipsing binaries
with component masses below 0.6 M$_{\odot}$ and well-determined
fundamental properties are known. However, even in these small numbers the
detailed comparison of the observed masses and radii with theoretical
predictions has revealed large disagreements. Current models seem to
predict radii of stars in the 0.4--0.8 M$_{\odot}$ range to be some
5--15\% smaller than observed. Given the high accuracy of the empirical
measurements (a few percent in both mass and radius), these differences
are highly significant. I review all the observational evidence on the
properties of M-type stars and discuss a possible scenario based on
stellar activity to explain the observed discrepancies.
\end{abstract}

\end{opening}           

\section{Introduction}

Most of the stars in the Galaxy have masses well below that of the Sun.  
In spite of the numerous population, detailed investigations of the
properties of low-mass stars has been often difficulted by their intrinsic
faintness. However, the observation and study of low-mass stars is now a
field in rapid development mostly because of the increasing number of deep
photometric surveys and the advent of powerful instrumentation able to
obtain spectroscopy of these faint stars. But also renewed interest arises
from one of the ``hot topics'' of this past decade: exoplanets. Very low
mass stars, brown dwarfs, and giant planets share many physical
characteristics and their study and modeling is often intimately related.

Efforts in the theoretical description of low mass stars have been intense
in recent years. Current stellar structure models of low mass stars have
reached a high level sophistication and maturity (e.g., Chabrier \&
Baraffe 2000). However, theoretical progress has not been matched by
observational developments because of the difficulty in obtaining accurate
determinations of the physical properties of low-mass stars. The
comparison of model predictions with observations is a central point. Only
by limiting the number of free parameters can a stringent test of stellar
models be carried out. Therefore, there is a strong need for stars with
well-determined properties such as masses, radii, effective temperatures,
metallicities and ages. Models will pass the test {\em only} if they are
able to reproduce {\em all} of the observed stellar properties.

The best source of such high-quality stellar properties is the analysis of
double-lined eclipsing binaries (EBs) in which the components are
detached. Optimum results are achieved when the system components are
similar (i.e., deep eclipses and two radial velocity sets). Unfortunately,
the number of known EBs with M-type components is small because of the
faintness of the stars and the often strong intrinsic variations due to
magnetic activity. To complement the dataset, in recent years it has
become possible to determine the radii of nearby M-type stars directly
from IR interferometry. The current precision does not match that of
eclipsing systems but the prospects are bright. Furthermore, planetary
transit research has also contributed to our database of low-mass stars.
Follow-up of OGLE transit candidates has uncovered a number of EBs with
F-G primaries and M-type secondaries.

In this paper I review the current data on masses and radii of low-mass
stars, including both EBs and stars with direct radius measurements, and
compare them with the predictions of stellar models. As already pointed
out by several authors (e.g., Torres \& Ribas 2002), a highly significant
discrepancy exists between observation and theory. Here I analyze possible
reasons for such discrepancy.

\section{Eclipsing M-type Systems}

Eclipsing binaries with similar components yield the stellar physical
properties potentially to an accuracy of 1--2\%. Such data have often
served as valuable benchmarks for the validation of structure and
evolution models. For two decades only two bona-fide EBs with M-type
components were known: The member of the Castor multiple system YY Gem
(Torres \& Ribas 2002), with components of spectral type M1~V, and CM Dra
(Lacy 1977; Metcalfe et al.  1996), composed of two M4.5~Ve stars.  These
were the only two M-type EBs that had been well studied until Delfosse et
al. (1999) reported the discovery of eclipses in the M3.5 star CU Cnc and
Ribas (2003) carried out accurate determinations of the components'
physical properties.  Very recently, three new M-type EBs have been
studied in detail. These are BW3 V38 (Maceroni \& Montalb\'an 2004),
TrES-Her0-07621 (Creevey et al. 2005), and GU Boo (L\'opez-Morales \&
Ribas 2005). Unfortunately, the quality of the available observations for
BW3 V38 and TrES-Her0-07621 does not permit high-accuracy determinations
of both masses and radii, which have uncertainties of up to 10--15\%. GU
Boo has well-determined physical properties that make it twin system of YY
Gem.  Table \ref{tab:ebs} gives the masses and radii of the components of
these M-type binaries. Also listed in the table are the components of two
eclipsing systems with masses below 0.8~M$_{\odot}$: The Hyades EB V818
Tau (Torres \& Ribas 2002) and RXJ0239.1-1028 (L\'opez-Morales et al., in
prep.). These K-type stars can be useful to better understand the results
of the comparison with stellar models.

\begin{table}[t]
\caption{Masses and radii of the components of double-lined EB
systems with masses below 0.8 M$_{\odot}$.}
\label{tab:ebs} 
\scriptsize
\begin{tabular}{lccl}
\hline
Name              & Mass (M$_{\odot}$)& Radius (R$_{\odot}$)& Ref. \\
\hline
V818 Tau B        & 0.7605$\pm$0.0062 &   0.768$\pm$0.010   & 1    \\
RXJ0239.1-1028 A  & 0.736$\pm$0.009   &   0.735$\pm$0.018   & -    \\
RXJ0239.1-1028 B  & 0.695$\pm$0.006   &   0.710$\pm$0.016   & -    \\
GU Boo A          & 0.610$\pm$0.007   &   0.623$\pm$0.016   & 2    \\
GU Boo B          & 0.599$\pm$0.006   &   0.620$\pm$0.020   & 2    \\
YY Gem AB         & 0.5992$\pm$0.0047 &   0.6191$\pm$0.0057 & 1    \\
TrES-Her0-07621 A & 0.493$\pm$0.003   &   0.453$\pm$0.060   & 3    \\
TrES-Her0-07621 B & 0.489$\pm$0.003   &   0.452$\pm$0.050   & 3    \\
BW3 V38 A         & 0.44$\pm$0.07     &   0.51$\pm$0.04     & 4    \\
BW3 V38 B         & 0.41$\pm$0.09     &   0.44$\pm$0.06     & 4    \\
CU Cnc A          & 0.4333$\pm$0.0017 &   0.4317$\pm$0.0052 & 5    \\
CU Cnc B          & 0.3890$\pm$0.0014 &   0.3908$\pm$0.0094 & 5    \\
CM Dra A          & 0.2307$\pm$0.0010 &   0.2516$\pm$0.0020 & 6,7  \\
CM Dra B          & 0.2136$\pm$0.0010 &   0.2347$\pm$0.0019 & 6,7  \\
\hline
\end{tabular}

Ref.: 1.- Torres \& Ribas (2002); 2.- L\'opez-Morales \& Ribas (2005); \\
3.- Creevey et al. (2005); 4.- Maceroni \& Montalb\'an (2004); \\
5.- Ribas (2003); 6.- Lacy (1977); 7.- Metcalfe et al. (1996). 
\end{table}

Stringent tests of stellar models can only be carried out if the chemical
compositions and ages of the EBs can be constrained. YY Gem and CU Cnc are
interesting cases because their kinematic properties indicate that they
belong to the so-called Castor moving group, with an age of approximately
$\sim$300~Myr and solar metallicity. Using also kinematic criteria, it
appears that GU Boo is an intermediate-age star in the galactic disk and
probably its metallicity is not far from the solar value. CM Dra is most
likely a Population II EB with sub-solar metallicity and an old age (of
the order of 10 Gyr). The ages and metallicities of BW3 V38 and
TrES-Her0-07621 have not been estimated. An important remark on the
significance of stellar ages is that M-type stars have long evolutionary
timescales once they have reached the main sequence. Thus, the only
relevant point to model comparisons is whether any of the studied EBs
could be pre-main sequence (i.e., an age $<$100 Myr). Available evidence
indicates that this is not the case.

\section{Other M-type Stars with Masses and Radii}

Besides double-lined EBs, other sources of masses and radii of low-mass
stars have emerged in recent years. Spectacular developments in
interferometry (such as the PTI or VLTI instruments) have made it possible
to resolve nearby M-type stars and determine their angular diameters with
uncertainties of just a few hundredths of a milliarcsecond. From those
measurements and trigonometric distances, determinations of stellar radii
can be carried out (Lane et al. 2001; S\'egransan et al. 2003). The
drawback of this technique is that the masses cannot be determined
directly (unless the resolved M-type star belongs to a visual binary) but
have to be inferred from calibrations. Fortunately, the empirical
mass-luminosity relationship in the infrared K band is well defined and
has little intrinsic scatter (Delfosse et al. 2000).  

Follow-up of OGLE
planetary transit candidates has uncovered a number of eclipsing systems
consisting of main sequence F-G stars with M dwarf companions (Bouchy et
al. 2005; Pont et al. 2005). Because of selection effects, their light
curves have shallow and flat-bottom eclipses corresponding to the transit
of the M-type star (the occultation not observable). Also, only the lines
of the F-G components are visible in the spectra due to the large
contrast. These restrictions imply that the masses and radii of the M-type
stars have to be determined through some assumptions (some of which are
model dependent). The resulting accuracies are in the range 5--20\%. The
masses and radii resulting from both interferometry and OGLE transit
follow-up are listed in Table \ref{tab:vbogle}.

\begin{table}[t]
\caption{Other low-mass stars with well-determined masses and radii.}
\label{tab:vbogle}
\scriptsize
\begin{tabular}{lccl}
\hline
Name         & Mass (M$_{\odot}$)  & Radius (R$_{\odot}$)& Ref.\\
\hline
OGLE-TR-114  & 0.82$\pm$0.08       &  0.72$\pm$0.09      & 1   \\
GJ 105A      & 0.790$\pm$0.039     &  0.708$\pm$0.050    & 2,3 \\
GJ 380       & 0.670$\pm$0.033     &  0.605$\pm$0.020    & 2,3 \\
GJ 205       & 0.631$\pm$0.031     &  0.702$\pm$0.063    & 3   \\
OGLE-TR-34   & 0.509$\pm$0.038     &  0.435$\pm$0.033    & 4   \\
GJ 887       & 0.503$\pm$0.025     &  0.491$\pm$0.014    & 3   \\
OGLE-TR-120  & 0.47$\pm$0.04       &  0.42$\pm$0.02      & 1   \\
GJ 15A       & 0.414$\pm$0.021     &  0.383$\pm$0.020    & 2,3 \\
GJ 411       & 0.403$\pm$0.020     &  0.393$\pm$0.008    & 2,3 \\
OGLE-TR-18   & 0.387$\pm$0.049     &  0.390$\pm$0.040    & 4   \\
OGLE-TR-6    & 0.359$\pm$0.025     &  0.393$\pm$0.018    & 4   \\
GJ 191       & 0.281$\pm$0.014     &  0.291$\pm$0.025    & 3   \\
OGLE-TR-7    & 0.281$\pm$0.029     &  0.282$\pm$0.013    & 4   \\
OGLE-TR-5    & 0.271$\pm$0.035     &  0.263$\pm$0.012    & 4   \\
OGLE-TR-78   & 0.243$\pm$0.015     &  0.240$\pm$0.013    & 1   \\
OGLE-TR-125  & 0.209$\pm$0.033     &  0.211$\pm$0.027    & 1   \\
GJ 699       & 0.158$\pm$0.008     &  0.196$\pm$0.008    & 2,3 \\
GJ 551       & 0.123$\pm$0.006     &  0.145$\pm$0.011    & 3   \\
OGLE-TR-106  & 0.116$\pm$0.021     &  0.181$\pm$0.013    & 1   \\
OGLE-TR-122  & 0.092$\pm$0.009     &  0.120$\pm$0.018    & 1   \\
\hline
\end{tabular}

Ref.: 1.- Pont et al. (2005); 2.- Lane et al. (2001); \\
3.- S\'egransan et al. (2003); 4.- Bouchy et al. (2005).
\end{table}

\section{Models versus Observations}

An obvious test of the performance of low-mass stellar models is to
compare the observational mass-luminosity diagram with theoretical
predictions. Most of the checks of state-of-the-art models using the
absolute magnitude in the V band have indicated good overall agreement but
significant scatter in the measurements. Further works (e.g., Delfosse et
al. 2000) have shown that such scatter is most likely caused by starspots
since the same mass-luminosity relationship is much better defined in the
infrared K band. From those tests, one may naively conclude that models
are successful at predicting the properties of low-mass stars. However,
this is a very restrictive comparison that uses only two of the several
independent properties that define a star.

The accurate masses and radii of the stars described above offer an
excellent opportunity to carry out critical tests to evaluate the
performance of low-mass stellar models. Such tests have been carried out
by a number of authors in the past (Popper 1997; Clausen et al. 1999;
Torres \& Ribas 2002; Ribas 2003), who have systematically pointed out a
(rather serious) discrepancy between the stellar radii predicted by theory
and the observations. Model calculations appear to underestimate stellar
radii by $\sim$10\%, which is a highly significant difference given the
observational uncertainties. Furthermore, the comparisons in some case
were made with virtually no free parameters since the ages and metal
contents of the stars could be constrained independently.

\begin{figure}
\begin{center}
\includegraphics[width=8.4cm]{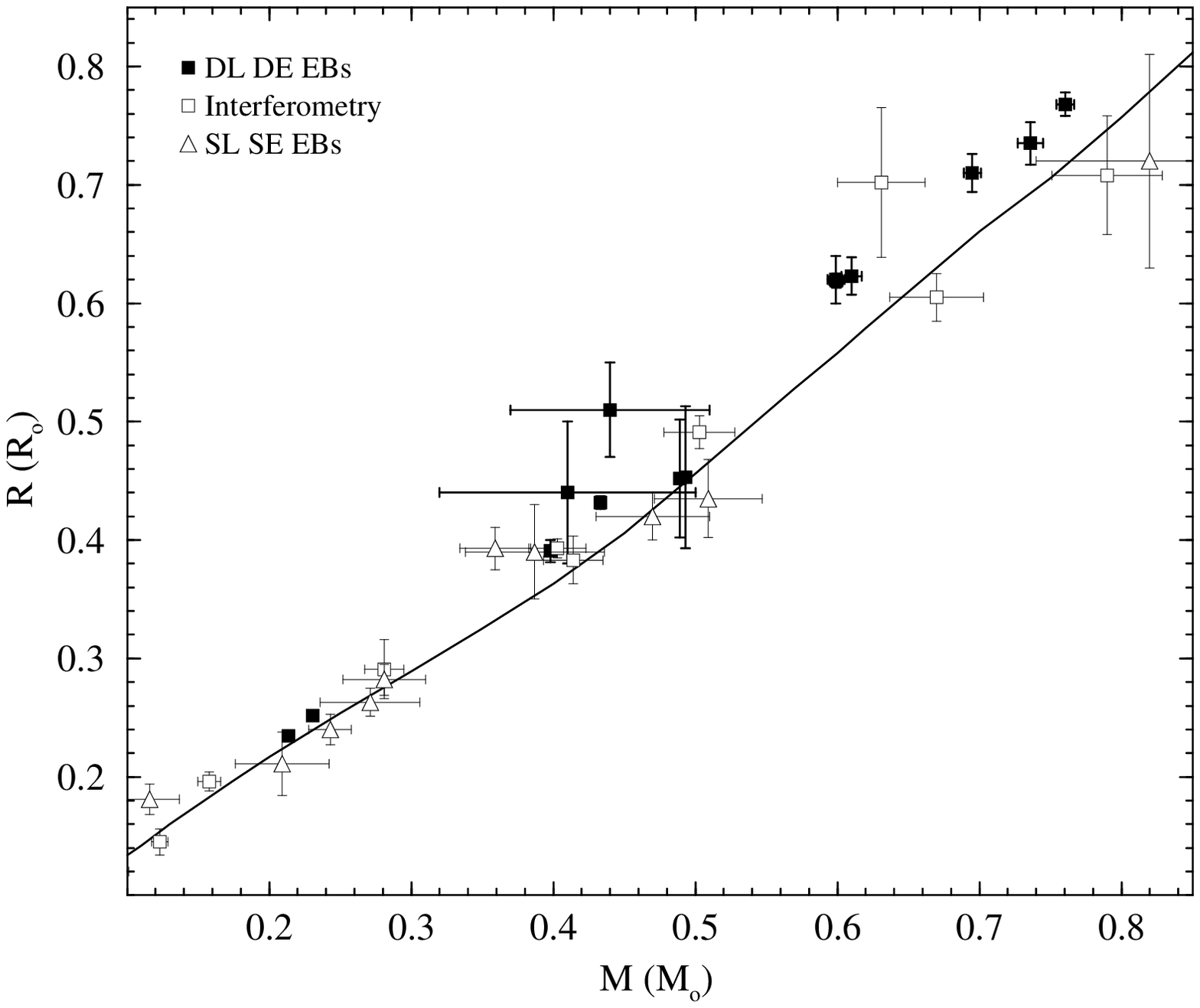}
\includegraphics[width=8.4cm]{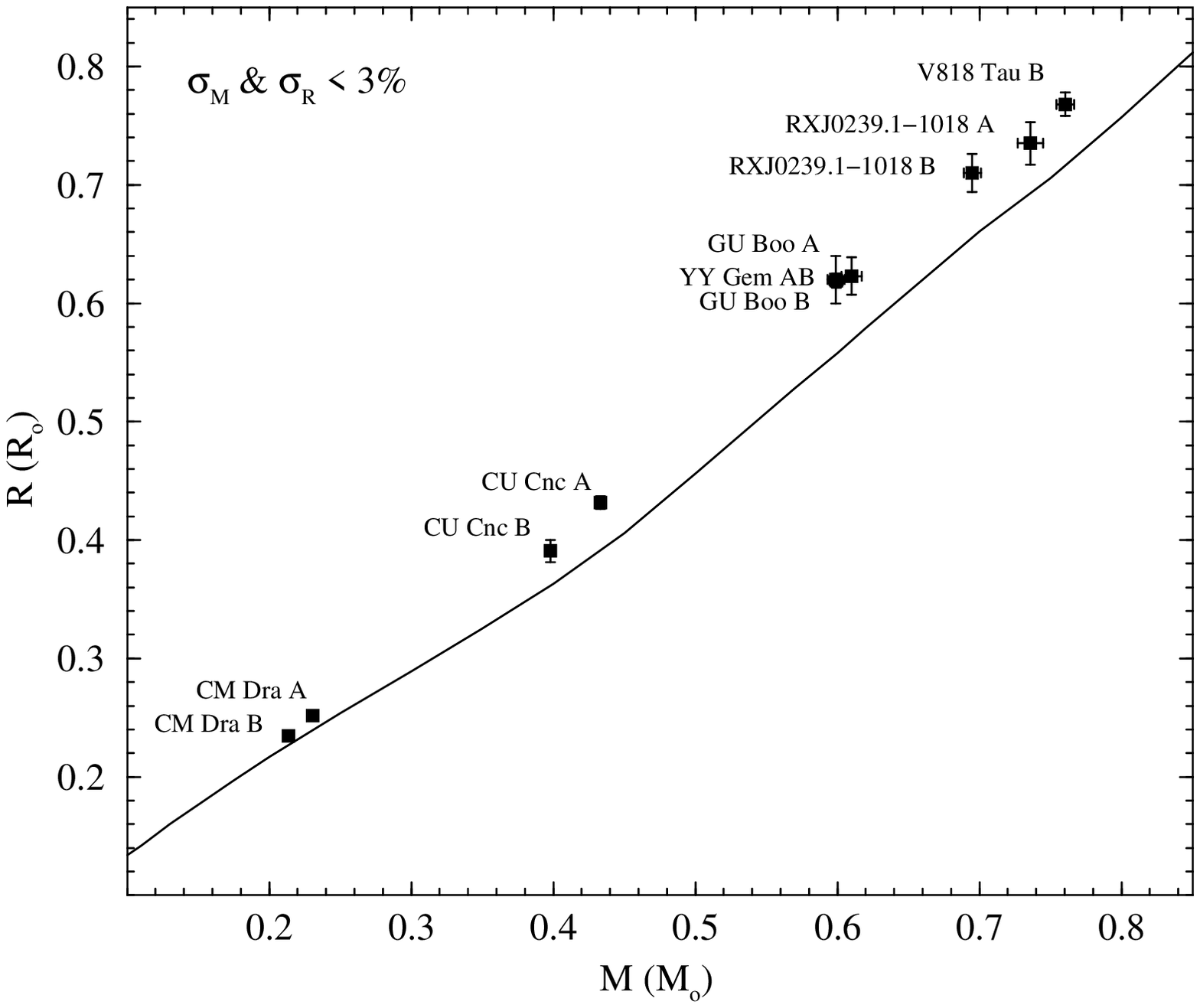}
\caption{{\bf Top:} Mass-radius plot for stars in the lower main sequence
with empirical determinations. The solid line represents a theoretical
300~Myr isochrone calculated with the Baraffe et al. (1998) models. {\bf
Bottom:} Same as above for those stars with determinations of masses and
radii better than 3\% (double-lined EBs).}
\label{fig:mr}
\end{center}
\end{figure}

With the extended stellar sample in this paper, the question of the
comparison between theory and observation can be revisited. Empirical
mass-radius diagrams are shown in Fig. \ref{fig:mr} showing both the
entire sample (top) and a subsample including those stars with masses and
radii determined to better than 3\% (bottom), which all happen to be EB
members. The line represents a 300 Myr isochrone (i.e., main sequence)
calculated with the models of Baraffe et al. (1998). Inspection of the top
panel shows two mass intervals with different characteristics. Stars with
masses below $\sim$0.30--0.35~M$_{\odot}$ seem to show small scatter and
good agreement with stellar models, while the more massive have larger
scatter and radii that tend to fall systematically above the theoretical line.
These distinct mass regimes are not well established yet but it is
tantalizing that the apparent division occurs near the limit between fully
convective stars and stars with radiative cores.

The high-accuracy sample in the bottom panel of Fig. \ref{fig:mr} leaves
no doubt that a significant discrepancy exists between models and
observations with regards to stellar radii. Other detailed comparisons
have also shown that the stellar effective temperatures appear to be
overestimated by $\sim$5\%. This, together with the good agreement in the
mass-luminosity plot, argues in favor of a scenario in which the stars
have larger radius and cooler temperature than predicted by models but
just in the right proportions to yield identical luminosities. A
$\sim$10\% radius undererestimation is compensated by a $\sim$5\%
temperature overestimation to yield identical luminosities. What would
explain such coincidence? The answer to this question is not clear yet,
but there are some hints pointing in certain directions.

Perhaps the first question to address is whether the EB sample used to
compare with models is representative of the low-mass star population.
These systems are all detached and should have evolved like single stars.
However, as members of close binaries (with periods less than 2.8 days)
the components have undergone tidal interactions forcing them to spin up
in orbital synchronism. The resulting high rotational velocities (10--60
km~s$^{-1}$) give rise to enhanced magnetic activity and thus to the
appearance of surface spots, emission lines and X-ray fluxes. As shown in
the work by Pizzolato et al. (2003), any M-type star with a rotation
period below 10 days will experience these phenomena at their peak
(saturated activity). It might be speculated that the larger radii and
lower temperatures could be a reflex of such enhanced activity. Perhaps
the significant spot areal coverage observed in these eclipsing systems
has the effect of lowering the overall photospheric temperature, which the
star compensates by increasing its radius to conserve the total radiative
flux. Thus, there may be a correlation between the radius and the activity
level of an M-type star. A similar conclusion of stellar activity causing
the discrepancy between models and theory was reached in the recent study
of Torres et al. (2005) for stars of higher masses
($\sim$0.9~M$_{\odot}$).

The sample we have used in our comparison may be representative of the
population of active M-type stars only. This does not diminish the
relevance of the discrepancy between models and observations. Low-mass
stars with ages younger than a few Gyr are very active because they are
generally fast rotators. Therefore, not only a star in a close binary
system but any active M-type star (e.g., in a stellar cluster) may have
its radius severely underestimated if computed from stellar models. A
definitive test of the magnetic activity hypothesis will have to wait for
further observational data. In particular, EBs with periods $\gtrsim$10
days (i.e., not synchronized) and components of visual binaries resolved
interferometrically should provide the necessary proof. Ongoing large
scale surveys and future space missions, such as COROT or Kepler, are
expected to increase the number of EBs significantly. If the activity
correlation is firmly established, it will be time for theory to catch up
by introducing magnetic activity in stellar evolution codes as a major
ingredient influencing the observable properties of low-mass stars.

\acknowledgements

I am grateful to E. Guinan, M. L\'opez-Morales and G. Torres for a number
of fruitful discussions. Support from the Spanish MCyT through a Ram\'on y
Cajal fellowship and grant AyA2003-07736 is acknowledged.

\end{article}
\end{document}